\documentclass[conference]{IEEEtran}
\IEEEoverridecommandlockouts
\usepackage{cite,amsmath,amssymb,amsfonts,algorithmic,graphicx,textcomp,xcolor,glossaries, cite, bm, bbm, hyperref, mathtools, comment, svg}

\def\BibTeX{{\rm B\kern-.05em{\sc i\kern-.025em b}\kern-.08em
    T\kern-.1667em\lower.7ex\hbox{E}\kern-.125emX}}
\newacronym{nn}{NN}{Neural Network}
\newacronym{dl}{DL}{Deep Learning}
\newacronym{cnn}{CNN}{Convolutional Neural Network}
\newacronym{wsq}{WSQ}{Wavelet Scalar Quantization}
\newacronym{wt}{WT}{Wavelet Transform}
\newacronym{msh}{MSH}{Mean and Scale Hyperprior}
\newacronym{mse}{MSE}{Mean Squared Error}
\newacronym{gdn}{GDN}{Generalized Divisive Normalization}
\newacronym{pca}{PCA}{Principal Component Analysis}
\newacronym{rd}{RD}{Rate-Distortion}
\newacronym{fp}{FP}{Fingerprint}
\begin{document}
\title{Effectiveness of learning-based image codecs on fingerprint storage}
%\thanks{\textbf{Restart PNRR etc etc}}
%
\makeatletter
\newcommand{\linebreakand}{%
  \end{@IEEEauthorhalign}
  \hfill\mbox{}\par
  \mbox{}\hfill\begin{@IEEEauthorhalign}
}
\makeatother
\author{
\IEEEauthorblockN{Daniele Mari}
\IEEEauthorblockA{
\textit{Università di Padova, Italy}\\
daniele.mari@dei.unipd.it}
\and
\IEEEauthorblockN{Saverio Cavasin}
\IEEEauthorblockA{
\textit{Università di Padova, Italy}\\
saverio.cavasin@phd.unipd.it}
\and
\IEEEauthorblockN{Simone Milani}
\IEEEauthorblockA{
\textit{Università di Padova, Italy}\\
simone.milani@dei.unipd.it}
\and
\IEEEauthorblockN{Mauro Conti}
\IEEEauthorblockA{
\textit{Università di Padova, Italy}\\
mauro.conti@unipd.it}
}
\newcommand{\sm}[1]{{\leavevmode\color{black} #1}}
\newcommand{\smd}[1]{{\leavevmode\color{black} #1}}
\newcommand{\smt}[1]{{\leavevmode\color{black} #1}}
\newcommand{\dm}[1]{{\leavevmode\color{black} #1}}

\maketitle
\begin{abstract}
The success of learning-based coding techniques and the development of learning-based image coding standards, such as JPEG-AI, point towards the adoption of such solutions in different fields, including the storage of biometric data, like fingerprints. However, the peculiar nature of learning-based compression artifacts poses several issues concerning their impact and effectiveness on extracting biometric features and landmarks, e.g., minutiae. This problem is utterly stressed by the fact that most models are trained on natural color images, whose characteristics are very different from usual biometric images, e.g, fingerprint or iris pictures. As a matter of fact, these issues are deemed to be accurately questioned and investigated, being such analysis still largely unexplored. 

This study represents the first investigation about the adaptability of learning-based image codecs in the storage of fingerprint images by measuring its impact on the extraction and characterization of minutiae.
Experimental results show that at a fixed rate point, learned solutions considerably outperform previous fingerprint coding standards, like JPEG2000, both in terms of distortion and minutiae preservation. Indeed, experimental results prove that the peculiarities of learned compression artifacts do not prevent automatic fingerprint identification (since minutiae types and locations are not significantly altered), nor do compromise image quality for human visual inspection (as they gain in terms of BD rate and PSNR of $47.8\%$ and $+3.97dB$ respectively).
\end{abstract}
\section{Introduction}
\label{sec:intro}
Fingerprints are one of the most widespread biometric identifiers, constituted by the raised papillary ridges that run across the skin’s surface. 
Their flow typically forms patterns that often present discontinuities due to breaks and deviations, known as \textit{minutiae} \cite{biometricsinstitute}. 
\sm{Such discontinuities are employed by \gls{fp} identification systems to align and compare the acquired image with users' templates stored in the database. As a matter of fact, preserving image quality while minimizing the required storage space proves to be crucial for database management and biometric data transmission. For these purposes, different image compression algorithms have been tested and adopted leading to the definition and standardization of guidelines and tests for codecs \cite{nist_fp}.}

\sm{These verifications are deemed by the fact that traditional image codecs introduce distortion like blurring (due to removal of high-pass components in signal reconstruction, e.g., JPEG2000 at low bit rates or deblocking filtering \cite{JPEG2000}), ringing (due to uncompensated frequency components erased by quantization) or blocking artifacts (depending on local independent block processing, e.g., JPEG or Intra coding for H.26x standards \cite{VVC}).}
\sm{As an example, it is possible to appreciate in the bottom right image in Fig.~\ref{fig:perceptual}, that was encoded with JPEG2000, that, especially in the lower half, ridges are blurred and several level $3$ details are removed.}
\sm{In relation to \gls{fp} coding, these result in the cancellation of minutiae and in the addition or cutting of ridges.}

Currently, the standard for \gls{fp} compression \sm{adopted} by FBI and NIST is \gls{wsq} \cite{brislawn1996fbi} which is based on the \gls{wt}. Additionally, also the widely known JPEG2000 \cite{JPEG2000} standard has been validated for \gls{fp} images and is commonly used \sm{whenever lossy} compression is required \cite{nist_fp}.

In recent years, learning-based algorithms for image coding have gained significant popularity due to their impressive performance gains \cite{balle2016end, theis2022lossy, balle2018variational, cheng2020learned, he2021checkerboard, jpeg-ai-cfp} compared to standard codecs \cite{JPEG2000, VVC}. This is mostly due to their inherent ability to learn entropy-efficient representations of the data and to learn and exploit strong priors. \sm{In learned image codecs, the input data are processed by a \gls{nn}, usually convolutional, whose filters might add fake minutiae or change their types and characteristics. 
As an example, some highly-optimized learned codecs \cite{mentzer2020high, yang2024lossy}, whose parameters were tuned to achieve high perceptual quality, might not be suitable for \gls{fp} coding since their inherently generative properties will likely affect minutiae's distribution. Moreover, \gls{fp} images are characterized by grayscale thin lines altered by sensor noise, smudges, finger pressure and alterations, dirt, and alien textures. These make \gls{fp}s very different from the traditional natural images on which codecs have been tuned upon, leading to reconstruction artifacts or rate inefficiencies.} 

\sm{Considering that most \gls{dl} codecs considerably outperform traditional schemes (including JPEG2000) and that the latest standardization efforts \cite{jpeg-ai-cfp} focus on learned solutions, it is very likely that, in the near future, the adoption of these techniques will be extended to different fields, such as biometric data, and, in particular, \gls{fp}.}
As an extra advantage, \glspl{nn} based approaches can be optimized for any type of loss function, so ad-hoc training strategies can be adopted in order to favor minutiae preservation.

\sm{Such reasoning and peculiarities suggest that the effectiveness of learned codecs for biometric images is to be investigated since nowadays it remains largely unexplored. To the best of our knowledge, this study is the first attempt aimed at evaluating the performance of such codecs on \gls{fp} compression and analyzing their impact on minutiae extraction and \gls{fp} characterization.}

\sm{Its} main contributions \sm{can be summarized as follows.}
\sm{\begin{enumerate}
\item{This work carries out the first thorough analysis of the impact of learned compression on the extraction of minutiae from coded \gls{fp} images is presented. More precisely, an evaluation of how learned compression alters, erases, or adds minutiae to the reconstructed \glspl{fp} is carried out.}
\item{The compression gains obtained by learned codecs with respect to previously standardized solutions, like JPEG2000 are measured and discussed.}
\item{It is proven that learned architectures present a valid upgrade for \gls{fp} coding both in terms of storage and image quality for both human inspection and automatic identification.}
\end{enumerate}}

The code for the paper is publicly available at https://github.com/Dan8991/Learning-based-fingerprint-coding.

\section{Background}
\label{sec:background}

\subsection{Fingerprints}

Fingerprints and biometric systems in general may operate either in \textit{verification} or \textit{identification} mode.

Several works have been exploring automated \gls{fp} identification and verification. In \cite{NURAALAM2021107387},  the authors combine features obtained from the Gabor filtering technique and machine learning techniques such as \glspl{cnn} and \gls{pca} to efficiently tackle identification.
Dalvi et al. \cite{dalvi2023deep} experiment with various \gls{dl} architectures for \gls{fp} enhancement, minutiae extraction, and \gls{fp} verification. 
Finally, the authors in \cite{fpclassification} propose the application of machine learning methods to develop \gls{fp} classification algorithms based on the singularity feature.

Additionally, multiple works have proposed algorithms that improve the quality of \glspl{fp} to enhance the performance in both tasks. For example, in \cite{fan} a U-net neural network with dilated convolutions is proposed for image denoising and inpainting.  
Yang et al.\cite{yang} propose a two-stage enhancement scheme that addresses the challenges posed by low-quality \gls{fp} inputs, such as cracks, scars, and poor ridge-valley contrast. Focusing on the matching requirements instead, Liu et al.\cite{liu} introduce a sparse coding-based orientation estimation algorithm for latent \glspl{fp}. 
Finally, in \cite{cavasin2024fingerprint} the authors investigated the membership and identity inference vulnerabilities of \gls{dl} models for \glspl{fp} images, showing that these approaches can lead to privacy issues.

\subsection{Image and Fingerprint compression}

Traditional lossy coding is often implemented via a transform coding framework, \sm{combined with adaptive predictions (e.g., spatial) and efficient entropy coding solutions (like context-adaptive arithmetic coding).}

\sm{Learned approaches also} follow this framework, with the encoder and decoder of an autoencoder serving as the analysis and synthesis transform. Building on this idea Ballé et al. \cite{balle2016end} propose to use an entropy model to estimate the probabilities of the latents to jointly optimize rate and distortion. Subsequent works have focused on improving both the analysis and synthesis transform and on increasing the representational power of the entropy model. For example, \cite{balle2018variational} proposes to use a hyperprior and to model the latents as a zero-mean Gaussian distribution. In \cite{minnen2018joint} two models are proposed, one that also estimates the mean of the Gaussian and one that additionally adds an autoregressive component to the entropy model.
Cheng et.al. \cite{cheng2020learned} introduce attention layers in the analysis and synthesis transform, and model the latents as Gaussian mixtures. Other works \cite{minnen2020channel, he2021checkerboard, he2022elic} improve the entropy model by making it channel autoregressive or by using a checkerboard pattern to reduce the number of autoregressive steps thus considerably reducing computation time by improving parallelism.
More recently the JPEG group has also started to standardize JPEG-AI \cite{jpeg-ai-cfp} an image codec that also uses an autoregressive entropy model but it also introduces many differences w.r.t. other codecs such as the separation of the Y and UV channels and ad-hoc training procedures. 

\begin{figure*}[ht]
    \centering
    \includegraphics[width=.8\textwidth]{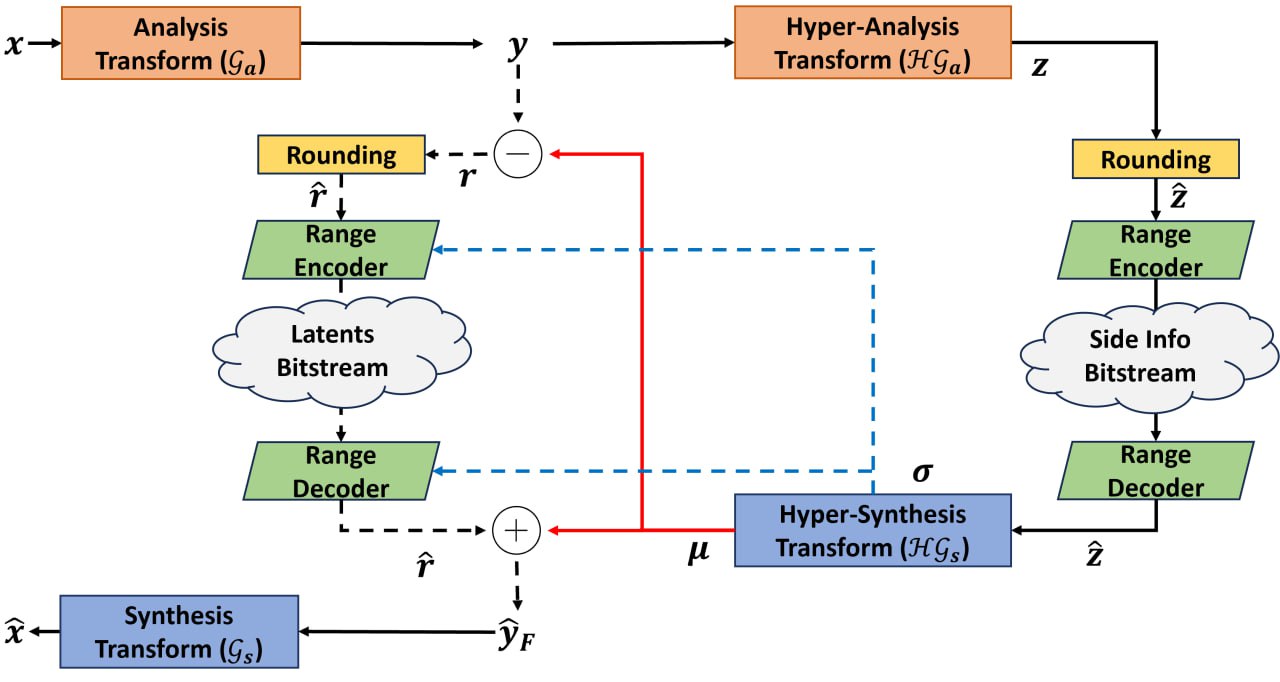}
    \caption{Learned image codec with a mean and scale hyperprior entropy model used as the main architecture in this work.}
    \label{fig:msh}
\end{figure*}

As previously mentioned the two most commonly used standards for \gls{fp} image coding are \gls{wsq} \cite{brislawn1996fbi} and JPEG2000 \cite{JPEG2000}. In JPEG 2000 the analysis transform is a discrete wavelet transform, then the obtained coefficients are quantized and entropy coded as explained above. Conversely, the synthesis transform consists of the inverse discrete wavelet transform. \gls{wsq} works similarly however it has a simpler entropy coding module.
Some other works that have tried to address \gls{fp} coding \cite{arthinovel} use K-SVD to obtain a sparse representation of the \gls{fp}. However, the research on the topic has been limited due to the versatility and high coding performance of JPEG2000.

\section{Method}
\label{sec:method}
\smd{In order to assess the performance of learned image codecs, the \gls{msh} model was selected since it is one core architecture that is adopted in multiple learned compression schemes and is highly performing despite having limited complexity. Since it does not present any autoregressive components}, the model is easily parallelizable thus reducing encoding and decoding time considerably w.r.t. its autoregressive counterparts. Tests were carried out both on a pre-trained model (trained on natural images) and on a new model trained from scratch specifically on \gls{fp} images called Finger-MSH. The architecture for Finger-MSH was changed minimally from the original to allow it to better process grayscale \gls{fp} images as will be explained in Sec~\ref{ref:figermsh}
\subsection{Mean and Scale Hyperprior}
The architecture of the model can be seen in Fig~\ref{fig:msh}. The codec can be used to encode an image by carrying out the following steps:
\begin{enumerate}
    \item the input image $x$ is fed to the analysis transform $\mathcal{G}_a$ to obtain the latent features $y = \mathcal{G}_a(x)$.
    \item the latent features $y$ are further compressed by the hyper-analysis transform $\mathcal{HG}_a$ to obtain the hyper-latents $z=\mathcal{HG}_a(y)$.
    \item the hyper-latents are then quantized $\hat{z} = \lceil z \rfloor$ and entropy coded according to the hyper-prior $P_{\hat{z}}(\hat{z})$, which is modeled according to a factorized prior model \cite{balle2018variational}, and stored/transmitted.
    \item the quantized hyper-latents $\hat{z}$ are fed to the hyper-synthesis transform to estimate the probability parameters $\mu, \sigma = \mathcal{HG}_s(\hat{z})$ for the latents $y$.
    \item the latents are then entropy coded according to the Gaussian distribution $\mathcal{N}(\mu, \sigma)$ and stored/transmitted.
\end{enumerate}
Conversely, when decoding:
\begin{enumerate}
    \item the quantized hyper-latents $\hat{z}$ are entropy-decoded using the hyper-prior $P_{\hat{z}} (\hat{z})$.
    \item the entropy parameters $\mu, \sigma$ are computed from $\hat{z}$ using the hyper-synthesis transform $\mu, \sigma = \mathcal{HG}_s(\hat{z})$
    \item the latents $\hat{y}$ are then entropy decoded according to the probability model  $\mathcal{N}(\mu, \sigma)$
    \item the image $\hat{x}$ is reconstructed by feeding the latents $\hat{y}$ to the synthesis transform $\mathcal{G}_s$ as $\hat{x} = \mathcal{G}_s(\hat{y})$
\end{enumerate}
During training the model is optimized in an end-to-end manner by using a differentiable approximation of quantization and by using the entropy of the latents as a proxy for the rate. In particular using the two probability models $P(\hat{z})$ and $P(\hat{y}|\hat{z})$ it is possible to approximate the entropy of the hyper-latents $\hat{z}$ 
\begin{equation}
    \mathcal{H}_{\hat{z}}(\hat{z}) \approx \sum_{\hat{z}_{i,j,k}} \frac{log_2(P(\hat{z}_{i,j,k}))}{|\hat{z}|}
\end{equation}
and latents $\hat{y}$  
\begin{equation}
    \mathcal{H}_{{\hat{y}}|{\hat{z}}}(\hat{y}) \approx \sum_{\hat{y}_{i,j,k}} \frac{log_2(P(\hat{y}_{i,j,k}|\hat{z}))}{|\hat{y}|}, 
\end{equation}
allowing to define the \gls{rd} loss
\begin{equation}
    \mathcal{L}(x) = \lambda\mathcal{D}(x, \hat{x}) + \mathcal{H}_{{\hat{y}}|{\hat{z}}}(\hat{y}) + \mathcal{H}_{\hat{z}}(\hat{z})
\end{equation}
where $\mathcal{D}$ is a distortion function such as \gls{mse}, $\lambda$ regulates the tradeoff between rate and distortion, and $\hat{y}, \hat{z}, \hat{x}$ are computed as previously explained using the main components of the network.

Generally, multiple models have to be trained, one for each required \gls{rd} tradeoff. This is necessary since the considered procedure allows optimizing the model only for a single \gls{rd} tradeoff specified by the $\lambda$ parameter. Some works, such as \cite{gao2022flexible} propose \smd{different training methods to enable different \gls{rd} points while using a single trained model. Such versatility is generally paid with a reduced  \gls{rd} performance.}

\subsection{Finger-MSH}
\label{ref:figermsh}
The implementation of the \gls{msh} model used in this paper is the one provided in the compressai \cite{begaint2020compressai} PyTorch library, whose architecture perfectly matches the one proposed in the original paper. However, some minor modifications had to be applied since the provided pre-trained models \smd{operate} on RGB natural images which are out of distribution w.r.t. grayscale \gls{fp} data. Furthermore, retraining on \gls{fp} images allows the model to learn proper priors thus leading to improved performance.
Additionally, the standard \gls{msh} model uses the \gls{gdn} \cite{balle2015density} activation function that is designed to improve the statistics of the activations of the network when processing natural images. In this case, \gls{gdn} was leading to unstable training, therefore, it was replaced by a standard LeakyReLU which is also less computationally demanding.
Finally, not all the $\lambda$ values suggested for training by the compressai library were used since after the sixth \gls{rd} point the quality of the reconstructed images reached saturation. The parameters for the convolutional layers in the analysis and synthesis transforms depend on the chosen $\lambda$ value, however, for consistency and for a fair comparison with the pre-trained model, the configuration was kept the same as the one in the original paper.

\section{Experiments and Results}
\label{sec:experiments}

The effectiveness of learned image codecs on fingerprints was tested both using the pre-trained models shared by compressai (labeled \gls{msh} in the figures) and on the adapted models (referred to as Finger-MSH) in order to see how much difference training on biometric data actually makes in the final performance. In this section, the training procedures for Finger-MSH and the experimental results are going to be presented. Since the pre-trained MSH codec works with RGB images, the fingerprints were coded by converting them to RGB before coding and then to grayscale again after coding.

\subsection{Dataset and training}
\smd{The dataset used to train and test Finger-MSH is the CASIA Fingerprint Dataset Version 5.0 (or CASIA-FingerprintV5) \cite{biometricsinstitute} collected by the Chinese Academy of Sciences’ Institute of Automation (CASIA). 
CASIA-FingerprintV5 contains 20.000 grayscale fingerprint images of 500 subjects (resolution $328\times356$). 
Each \smd{subject} contributed 40 fingerprint images, 5 per finger excluding the little fingers, and each acquisition presents significant quality differences  (partiality, cuts, dirt, blurriness).}

\begin{figure}
    \centering
    \includegraphics[width=.45\columnwidth]{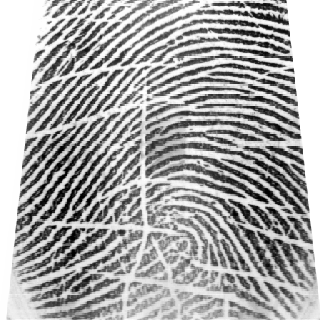}
    \includegraphics[width=.45\columnwidth]{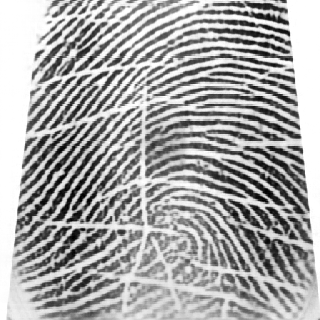}
    
    \vspace*{.3em}
    
    \includegraphics[width=.45\columnwidth]{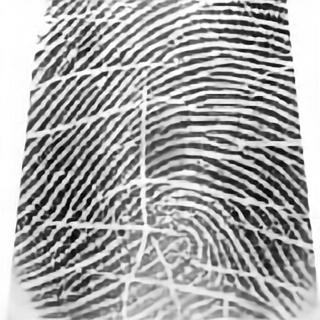}
    \includegraphics[width=.45\columnwidth]{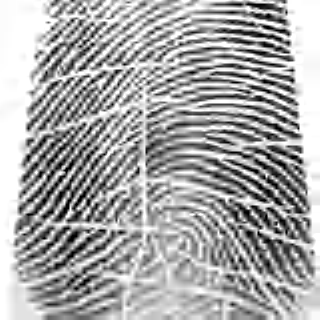}
    \caption{Original at $\sim$ 74kB (top-left), compressed with Finger-MSH at $\sim$ 4kB, PSNR=24.84dB (top-right), compressed with MSH at $\sim$ 5kB, PSNR=23.03dB (bottom-left), compressed with JPEG2000 at $\sim$ 5kB, PSNR=17.79 dB (bottom-right)}
    \label{fig:perceptual}
    \vspace{-.5em}
\end{figure}

\begin{figure}
    \centering
    \includegraphics[width=.9\columnwidth]{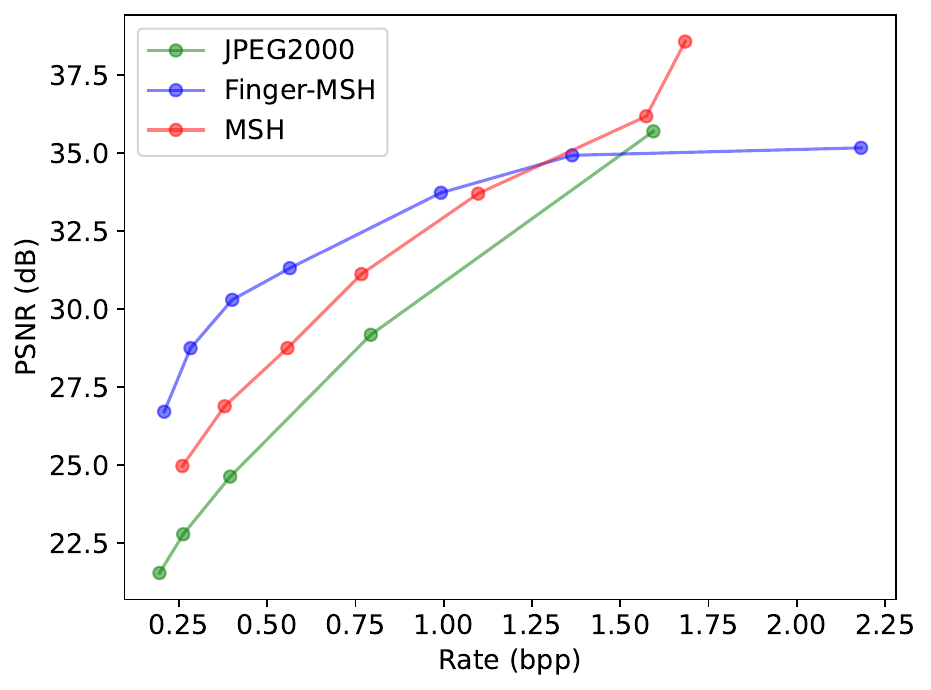}
    \includegraphics[width=.9\columnwidth]{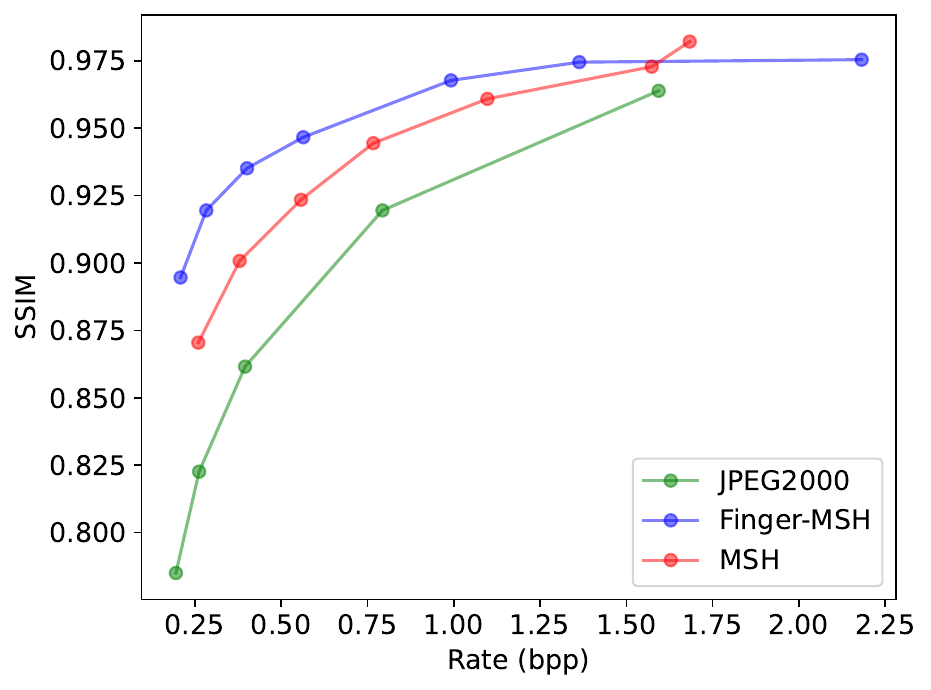}
    \caption{RD curves for PSNR (top) and SSIM (bottom)}
    \label{fig:rd}
    \vspace{-1em}
\end{figure}

\begin{figure*}
    \centering
    \includegraphics[width=.32\textwidth]{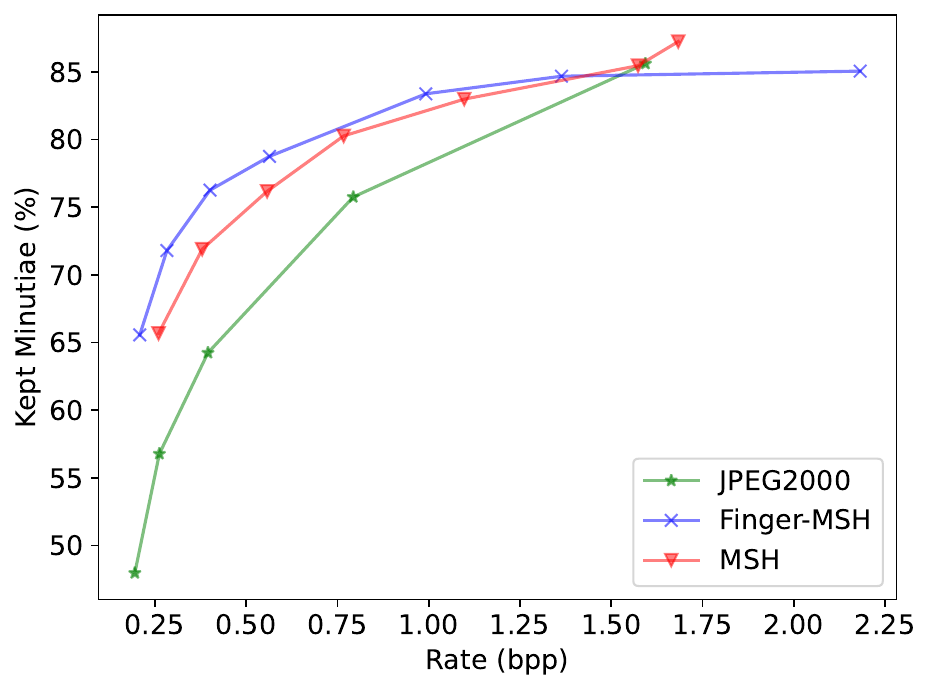}
    \includegraphics[width=.32\textwidth]{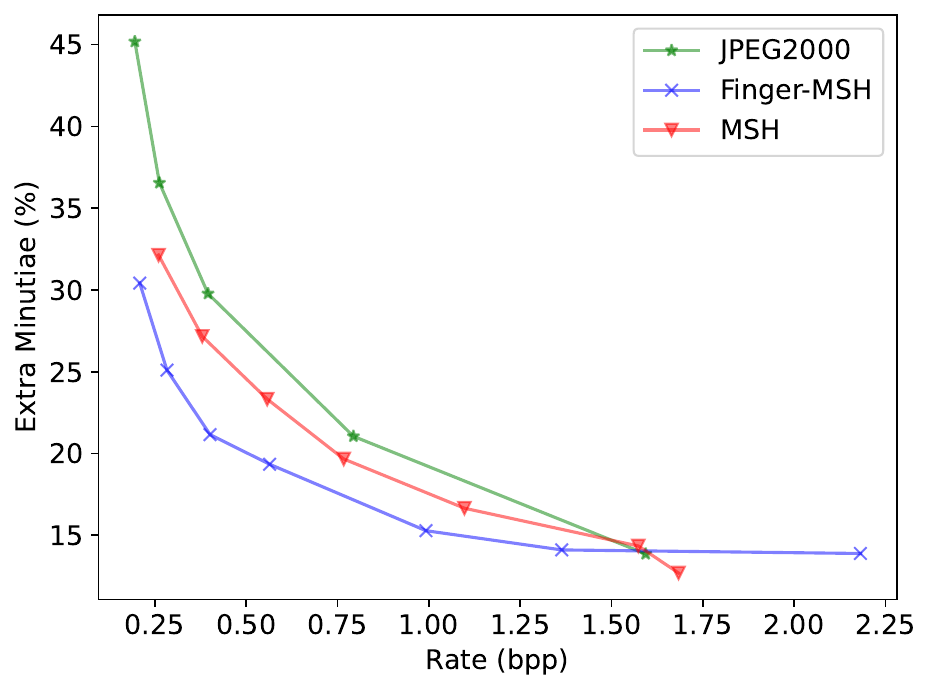}
    \includegraphics[width=.32\textwidth]{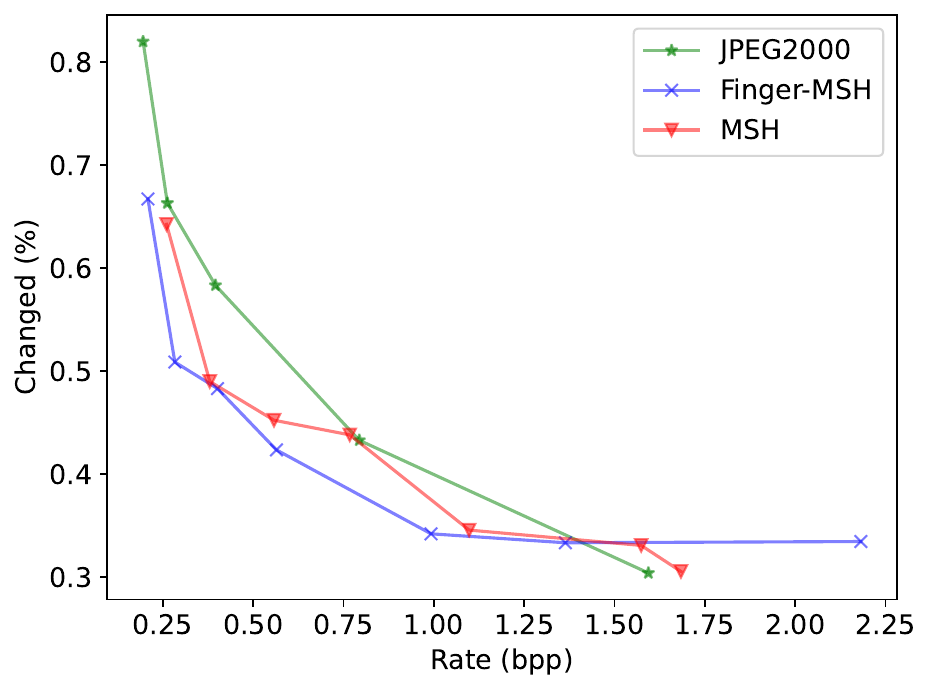}
    \includegraphics[width=.32\textwidth]{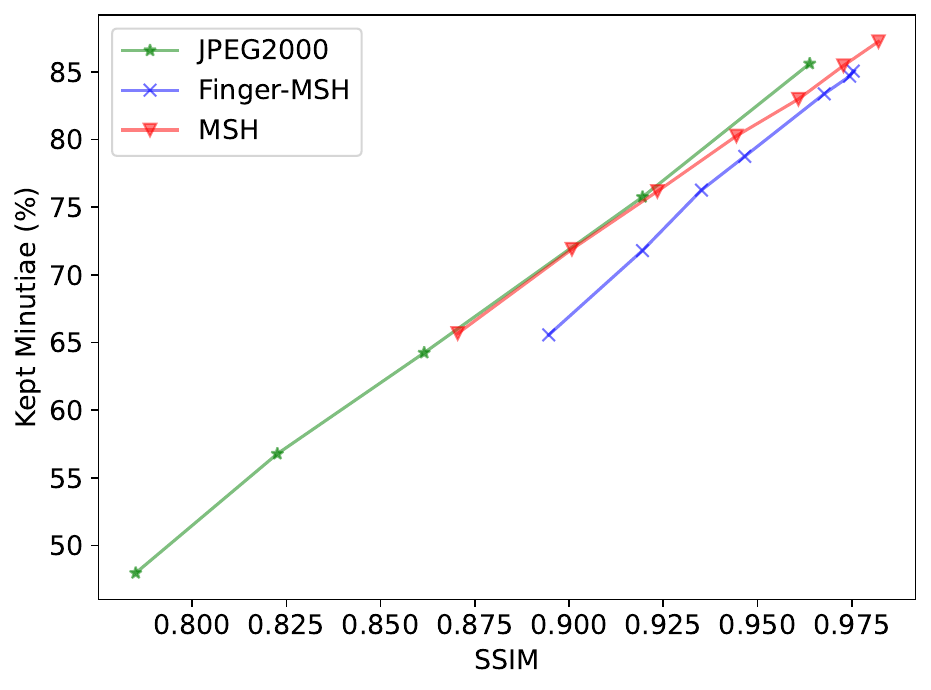}
    \includegraphics[width=.32\textwidth]{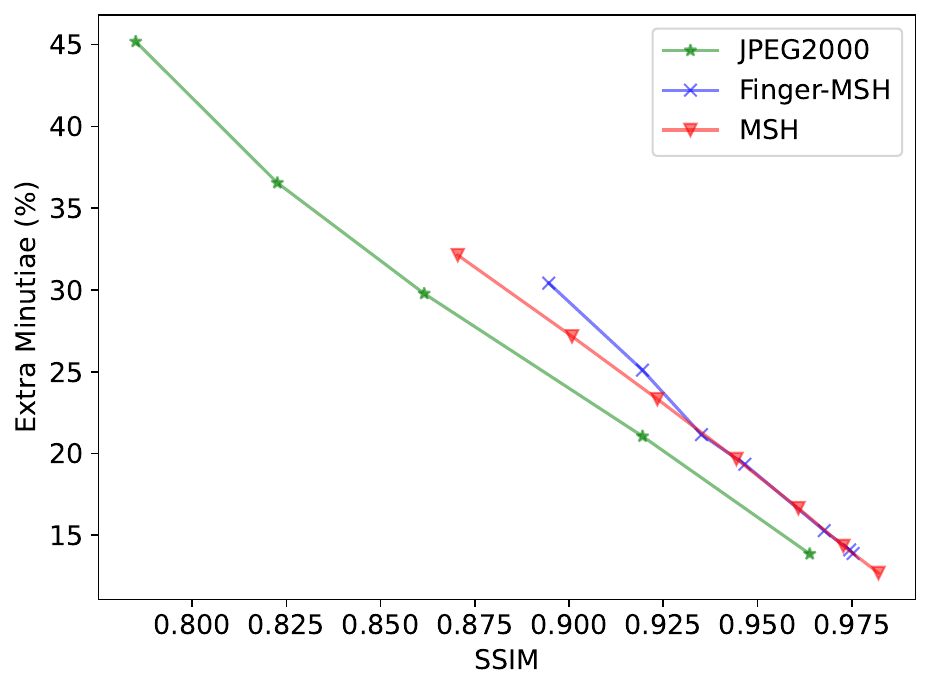}
    \includegraphics[width=.32\textwidth]{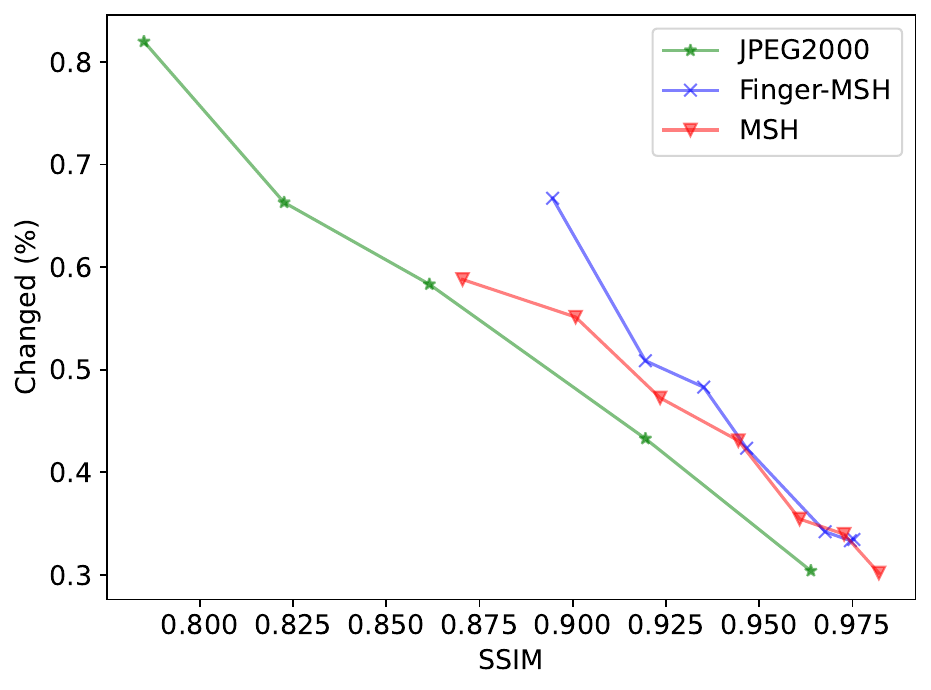}
    \caption{Correctly kept minutiae (left column), extra introduced minutiae (central column), and minutiae that changed type (right column) as a function of rate (first row) and \dm{SSIM} (second row).}
    \label{fig:minutiae}
    \vspace{-1em}
\end{figure*}

The dataset was split into training, validation, and test sets with a 60/20/20\% split by keeping the first 300 users for training, the next 100 for validation, and the last 100 for testing. Since the \gls{msh} and Finger-MSH models can only process images whose height and width are a multiple of 64 the images were center cropped to a shape of $320\times320$. Padding can also be used to obtain the proper image shape, however, in this case, cropping was chosen since the fingerprints are generally centered in the image so it did not lead to any loss of information.

Following the compressai configurations the tradeoff values used for training are $\lambda \in \{0.0018, 0.0035, 0.0067, 0.013, 0.025, 0.0483, 0.0932\}$. Each model is trained for at most 1000 epochs using Adam optimizer \cite{kingma2014adam} with a learning rate starting from 0.0001 decayed by a factor of 2 every 20 epochs without improvement. The training was stopped whenever the learning rate reached a value of $5\times10^{-6}$. The images were encoded with JPEG2000 using compression ratio parameters $40, 30, 20, 10, 5$.

\subsection{Evaluation metrics}
Finger-MSH and MSH were tested against JPEG2000 and compared in terms of \gls{rd} performance. As a next step, the effect of compression on the number of detected minutiae was analyzed. In particular, the compressed images were enhanced using the \textit{fingerprint\_enhancer} python library that uses unoriented Gabor filters to clean the fingerprint images similarly to what was proposed in \cite{hong1998fingerprint}. %This helps increase the minutiae detection accuracy. 
Then, the terminations and bifurcations were extracted using the \textit{fingerprint\_feature\_extractor} library both in the original and in the compressed images and matched using a distance-based algorithm with a threshold of 3. At this point, three metrics were computed i.e.:
\begin{itemize}
    \item kept terminations/bifurcations: minutiae that were correctly preserved after compression
    \item extra terminations/bifurcations: minutiae that were not present in the original image but that were detected after compression
    \item \smd{changed terminations/bifurcations: these are minutiae that changed their types after compression (e.g. termination to bifurcation or bifurcation to termination).}
\end{itemize}
These quantities were plotted firstly as a function of the rate to understand which of the two approaches is more convenient in a real-case scenario. This is particularly useful also because the learned network is at a slight advantage when compressing this type of data since the outer part of the image (the sensor) is always the same in the CASIA dataset, i.e. the network can always reconstruct it perfectly thus slightly decreasing the average distortion. For this reason, using a measure of distortion that only depends on the fingerprint provides a fairer comparison. Additionally, the metrics on the minutiae were plotted as a function of \dm{SSIM} to understand if at comparable qualities learned methods tend to have greater or smaller effect on the minutiae w.r.t. JPEG2000.

All the proposed metrics are computed for all the 4000 images of the test set and averaged to obtain the final results.

\subsection{Results}
The performance of Finger-MSH and MSH will be validated perceptually, in terms of \gls{rd} curves and in terms of the metrics on the minutiae computed above. In particular in Fig~\ref{fig:perceptual} it is possible to see that even at a lower rate the learned codecs produce a much sharper image that is more perceptually similar to the original one with far less blurring. Additionally also from the RD-curves (see Fig~\ref{fig:rd}) the learned models show that they both lead to considerably lower distortion for comparable rates (this holds both for PSNR and SSIM). In particular, w.r.t. JPEG2000, Finger-MSH leads to $47.8\%$ BD rate savings and $3.97dB$ BD PSNR gains, while MSH saves on average $24.8\%$ in terms of rate and gains $1.97dB$ in terms of PSNR. When comparing Finger-MSH and MSH it is possible to see that training on fingerprints gives a big advantage at lower rates, however at the highest ones it seems that \gls{msh} can achieve higher fidelity while Finger-MSH saturates earlier. This is likely due to the amount of data used for training. While Finger-MSH was trained on 12000 images, MSH was trained on more than an order of magnitude more images allowing it to achieve reach lower distortions. The saturation of the distortion metric motivates the chosen range of $\lambda$ values for Finger-MSH.

\smt{As for the previously defined minutiae metrics, it is possible to notice that learned codecs considerably outperform JPEG2000  at a given bit rate (see Fig~\ref{fig:minutiae} top row) thanks to their RD performance}. 
\smt{The bottom row of Fig~\ref{fig:minutiae} reports the preservation metrics for a given SSIM value (equalizing objective quality metric): in this case JPEG2000 seems to perform better although it requires a significantly higher bit rate cost. This effect can be mitigated by including more fingerprint samples \cite{app131810000} in the training set (justified by the performance of MSH over Finger-MSH ) and including some minutiae preservation metric in the loss.}
\section{Conclusions}
\label{sec:conclusions}
\smd{This paper is a first attempt to discuss the adoption of learned compression on fingerprint images and evaluate the impact of the related coding artifacts on the authentication and identification procedures.}

\smd{As it has been shown on natural images, the compressed representations learned by \gls{nn}-based codecs are more entropically efficient than the ones obtained by traditional fingerprint-related codecs, saving approximately 50\% of the bitrate on average}. This is likely due to the fact that the network can learn some strong priors that can be improved by training the model on task-specific data (in this case fingerprints). 
Additionally, the coded images yield better minutiae preservation at comparable rates w.r.t. JPEG2000 \dm{suggesting that NN-based codecs are valuable solutions for the storage of biometric data thus enjoying the RD gains they bring. Nevertheless,} \smt{since codecs are trained on RD functions related to standard image compression, at a given quality value the added artifacts have a slightly worsening effect on the minutiae preservation w.r.t. JPEG 2000.}
Finally, experimental results show that, although trained on out-of-distribution images, pretrained learned image codecs outperform standard ones. For this reason, the adoption of codecs such as JPEG-AI, even without retraining, is likely to provide considerable \gls{rd} gains.
\smt{Motivated by the results obtained equalizing the reconstruction quality and by the possibility of optimizing \gls{nn}s on multiple distortion metrics,} future research will extend the training  \dm{losses including} minutiae's preservation metric to maximize the matching performance; \smt{ moreover, we are going to assess the performance of learned codecs on other biometric identifiers.}
\section*{Acknowledgments}
This work was partially supported by the European Union under the Italian National Recovery and Resilience Plan (NRRP) of NextGenerationEU, with a partnership on “Telecommunications of the Future” (PE00000001 - program “RESTART”). This study was also carried out within the Future Artificial Intelligence Research (FAIR) and received funding from the European Union Next-GenerationEU (PNRR - Piano Nazionale di Ripresa e Resilienza - Missione 4, Componente 2, Investimento 1.3 - D.D. 1555 11/10/2022, PE00000013). This manuscript reflects only the authors’ views and opinions, neither the EU nor the European Commission can be considered responsible for them. Daniele Mari's activities were supported by Fondazione CaRiPaRo under the grants “Dottorati di Ricerca” 2021/2022.
\bibliographystyle{IEEEtran}
\bibliography{papers}

% Generated by IEEEtran.bst, version: 1.12 (2007/01/11)
\begin{thebibliography}{10}
\providecommand{\url}[1]{#1}
\csname url@samestyle\endcsname
\providecommand{\newblock}{\relax}
\providecommand{\bibinfo}[2]{#2}
\providecommand{\BIBentrySTDinterwordspacing}{\spaceskip=0pt\relax}
\providecommand{\BIBentryALTinterwordstretchfactor}{4}
\providecommand{\BIBentryALTinterwordspacing}{\spaceskip=\fontdimen2\font plus
\BIBentryALTinterwordstretchfactor\fontdimen3\font minus
  \fontdimen4\font\relax}
\providecommand{\BIBforeignlanguage}[2]{{%
\expandafter\ifx\csname l@#1\endcsname\relax
\typeout{** WARNING: IEEEtran.bst: No hyphenation pattern has been}%
\typeout{** loaded for the language `#1'. Using the pattern for}%
\typeout{** the default language instead.}%
\else
\language=\csname l@#1\endcsname
\fi
#2}}
\providecommand{\BIBdecl}{\relax}
\BIBdecl

\bibitem{biometricsinstitute}
\BIBentryALTinterwordspacing
B.~Institute. Types of biometrics - fingerprint: Key considerations. [Online].
  Available:
  \url{https://www.biometricsinstitute.org/types-of-biometrics-fingerprint-key-considerations/}
\BIBentrySTDinterwordspacing

\bibitem{nist_fp}
\BIBentryALTinterwordspacing
S.~Orandi, J.~Libert, M.~Garris, J.~Grantham, and F.~Byers,
  ``\BIBforeignlanguage{en}{Jpeg 2000 codec certification guidance for 1000 ppi
  fingerprint friction ridge imagery},'' may 2021. [Online]. Available:
  \url{https://nvlpubs.nist.gov/nistpubs/SpecialPublications/NIST.SP.500-300-upd.pdf}
\BIBentrySTDinterwordspacing

\bibitem{JPEG2000}
\emph{{JPEG 2000 Part I: Final Draft International Standard}}, 2000.

\bibitem{VVC}
B.~Bross, Y.-K. Wang, Y.~Ye, S.~Liu, J.~Chen, G.~J. Sullivan, and J.-R. Ohm,
  ``Overview of the versatile video coding (vvc) standard and its
  applications,'' \emph{IEEE Transactions on Circuits and Systems for Video
  Technology}, vol.~31, no.~10, pp. 3736--3764, 2021.

\bibitem{brislawn1996fbi}
C.~M. Brislawn, J.~N. Bradley, R.~J. Onyshczak, and T.~Hopper, ``Fbi
  compression standard for digitized fingerprint images,'' in
  \emph{Applications of digital image processing XIX}, vol. 2847.\hskip 1em
  plus 0.5em minus 0.4em\relax Spie, 1996, pp. 344--355.

\bibitem{balle2016end}
J.~Ball{\'e}, V.~Laparra, and E.~P. Simoncelli, ``End-to-end optimized image
  compression,'' \emph{arXiv preprint arXiv:1611.01704}, 2016.

\bibitem{theis2022lossy}
L.~Theis, W.~Shi, A.~Cunningham, and F.~Husz{\'a}r, ``Lossy image compression
  with compressive autoencoders,'' in \emph{International conference on
  learning representations}, 2022.

\bibitem{balle2018variational}
J.~Ball{\'e}, D.~Minnen, S.~Singh, S.~J. Hwang, and N.~Johnston, ``Variational
  image compression with a scale hyperprior,'' \emph{arXiv preprint
  arXiv:1802.01436}, 2018.

\bibitem{cheng2020learned}
Z.~Cheng, H.~Sun, M.~Takeuchi, and J.~Katto, ``Learned image compression with
  discretized gaussian mixture likelihoods and attention modules,'' in
  \emph{Proceedings of the IEEE/CVF conference on computer vision and pattern
  recognition}, 2020, pp. 7939--7948.

\bibitem{he2021checkerboard}
D.~He, Y.~Zheng, B.~Sun, Y.~Wang, and H.~Qin, ``Checkerboard context model for
  efficient learned image compression,'' in \emph{Proceedings of the IEEE/CVF
  Conference on Computer Vision and Pattern Recognition}, 2021, pp.
  14\,771--14\,780.

\bibitem{jpeg-ai-cfp}
\emph{{Final Call for Proposals for JPEG AI}}, 2022.

\bibitem{mentzer2020high}
F.~Mentzer, G.~D. Toderici, M.~Tschannen, and E.~Agustsson, ``High-fidelity
  generative image compression,'' \emph{Advances in Neural Information
  Processing Systems}, vol.~33, pp. 11\,913--11\,924, 2020.

\bibitem{yang2024lossy}
R.~Yang and S.~Mandt, ``Lossy image compression with conditional diffusion
  models,'' \emph{Advances in Neural Information Processing Systems}, vol.~36,
  2024.

\bibitem{NURAALAM2021107387}
\BIBentryALTinterwordspacing
Nur-A-Alam, M.~Ahsan, M.~Based, J.~Haider, and M.~Kowalski, ``An intelligent
  system for automatic fingerprint identification using feature fusion by gabor
  filter and deep learning,'' \emph{Computers and Electrical Engineering},
  vol.~95, p. 107387, 2021. [Online]. Available:
  \url{https://www.sciencedirect.com/science/article/pii/S0045790621003554}
\BIBentrySTDinterwordspacing

\bibitem{dalvi2023deep}
N.~Dalvi and V.~V. Pham, ``Deep learning approaches for fingerprint
  verification,'' 2023.

\bibitem{fpclassification}
N.~Thu~Huong and N.~The~Long, ``Fingerprints classification through image
  analysis and machine learning method,'' \emph{Algorithms}, vol.~12, p. 241,
  11 2019.

\bibitem{fan}
W.~Fan, ``Deep encoder-decoder neural network for fingerprint image denoising
  and inpainting,'' 05 2020.

\bibitem{yang}
J.~Yang, N.~Xiong, and A.~V. Vasilakos, ``Two-stage enhancement scheme for
  low-quality fingerprint images by learning from the images,'' \emph{IEEE
  Transactions on Human-Machine Systems}, vol.~43, no.~2, pp. 235--248, 2013.

\bibitem{liu}
S.~Liu, M.~Liu, and Z.~Yang, ``Sparse coding based orientation estimation for
  latent fingerprints,'' \emph{Pattern Recognition}, vol.~67, 02 2017.

\bibitem{cavasin2024fingerprint}
S.~Cavasin, D.~Mari, S.~Milani, and M.~Conti, ``Fingerprint membership and
  identity inference against generative adversarial networks,'' 2024.

\bibitem{minnen2018joint}
D.~Minnen, J.~Ball{\'e}, and G.~D. Toderici, ``Joint autoregressive and
  hierarchical priors for learned image compression,'' \emph{Advances in neural
  information processing systems}, vol.~31, 2018.

\bibitem{minnen2020channel}
D.~Minnen and S.~Singh, ``Channel-wise autoregressive entropy models for
  learned image compression,'' in \emph{2020 IEEE International Conference on
  Image Processing (ICIP)}.\hskip 1em plus 0.5em minus 0.4em\relax IEEE, 2020,
  pp. 3339--3343.

\bibitem{he2022elic}
D.~He, Z.~Yang, W.~Peng, R.~Ma, H.~Qin, and Y.~Wang, ``Elic: Efficient learned
  image compression with unevenly grouped space-channel contextual adaptive
  coding,'' in \emph{Proceedings of the IEEE/CVF Conference on Computer Vision
  and Pattern Recognition}, 2022, pp. 5718--5727.

\bibitem{arthinovel}
S.~Arthi and J.~Stanly~Jayaprakash, ``A novel approach for fingerprint sparse
  coding analysis using k-svd learning technique.''

\bibitem{gao2022flexible}
C.~Gao, T.~Xu, D.~He, Y.~Wang, and H.~Qin, ``Flexible neural image compression
  via code editing,'' \emph{Advances in Neural Information Processing Systems},
  vol.~35, pp. 12\,184--12\,196, 2022.

\bibitem{begaint2020compressai}
J.~B{\'e}gaint, F.~Racap{\'e}, S.~Feltman, and A.~Pushparaja, ``Compressai: a
  pytorch library and evaluation platform for end-to-end compression
  research,'' \emph{arXiv preprint arXiv:2011.03029}, 2020.

\bibitem{balle2015density}
J.~Ball{\'e}, V.~Laparra, and E.~P. Simoncelli, ``Density modeling of images
  using a generalized normalization transformation,'' \emph{arXiv preprint
  arXiv:1511.06281}, 2015.

\bibitem{kingma2014adam}
D.~P. Kingma and J.~Ba, ``Adam: A method for stochastic optimization,''
  \emph{arXiv preprint arXiv:1412.6980}, 2014.

\bibitem{hong1998fingerprint}
L.~Hong, Y.~Wan, and A.~Jain, ``Fingerprint image enhancement: algorithm and
  performance evaluation,'' \emph{IEEE transactions on pattern analysis and
  machine intelligence}, vol.~20, no.~8, pp. 777--789, 1998.

\bibitem{app131810000}
\BIBentryALTinterwordspacing
A.~Makrushin, V.~S. Mannam, and J.~Dittmann, ``Privacy-friendly datasets of
  synthetic fingerprints for evaluation of biometric algorithms,''
  \emph{Applied Sciences}, vol.~13, no.~18, 2023. [Online]. Available:
  \url{https://www.mdpi.com/2076-3417/13/18/10000}
\BIBentrySTDinterwordspacing

\end{thebibliography}
\end{document}